\documentclass[10pt]{article}
\usepackage{times}
\usepackage{amsmath}
\usepackage{amssymb}
\usepackage{stmaryrd}
\usepackage[english]{babel}
\usepackage{supertabular }
\usepackage{euscript}
\usepackage{graphicx}
\usepackage{tabularx}

\newcommand{\syntaxdef}{::=}
\newcommand{\procdef}{\displaystyle \mathop{=}^{\mbox{\scriptsize def}}}
\newcommand{\nattype}{\mbox{Nat}}
\newcommand{\qubittype}{\mbox{Qubit}}
\newcommand{\qvar}[1]{#1:\qubittype}
\newcommand{\cvar}[1]{#1:\nattype}
\newcommand{\declvar}[2]{#1: #2}
\newcommand{\findeclvar}{\pmb\ .\ }
\newcommand{\debutbloc}{\pmb [\ }
\newcommand{\finblocse}{\pmb ]}
\newcommand{\finbloc}{\ \pmb ]}
\newcommand{\prefix}{.}
\newcommand{\seq}{\ ;}
\newcommand{\para}{\parallel}
\newcommand{\nondet}{+}
\newcommand{\cond}[2]{#1 \rightarrow #2}
\newcommand{\restrict}[1]{\backslash#1\ }
\newcommand{\restrictg}[1]{\backslash\{#1\}\ }
\newcommand{\envoi}[1]{\ !#1\  }
\newcommand{\recep}[1]{\ ?#1\  }
\newcommand{\term}{\mbox{\it end}}
\newcommand{\stopproc}{\mbox{\it nil}}
\newcommand{\mathoparrow}{\mathop{\longrightarrow}}

\newcommand{\atransition}[1]
{\hspace{10pt}\underrightarrow{\hspace{10pt}#1\hspace{10pt}}\hspace{10pt}}

\newcommand{\petiteatransition}[1]{\mbox{$\displaystyle\ \mathoparrow^{#1}\ $}}
\newcommand{\tautransition}
{\hspace{10pt}\underrightarrow{\hspace{10pt}\tau\hspace{10pt}}\hspace{10pt}}

\newcommand{\petitetautransition}{\petiteatransition{\tau}}
\newcommand{\dtransition}
{\hspace{10pt}\underrightarrow{\hspace{10pt}\delta\hspace{10pt}}\hspace{10pt}}

\newcommand{\petitedtransition}{\petiteatransition{\delta}}
\newcommand{\ptransition}[1]{\hspace{10pt}\longrightarrow_{#1}\hspace{10pt}}
\newcommand{\ptransitionse}[1]{\longrightarrow_{#1}}

\newcommand{\ket}[1]{|#1\rangle}
\newcommand{\bra}[1]{\langle#1|}
\newcommand{\contexte}[4]{<#1, #2 = #3, #4 >}
\newcommand{\scontexte}[4]{/<#1, #2 = #3, #4 >}
\newcommand{\spcontexte}[1]{/ #1}
\newcommand{\contexteqcq}{<s, q = \rho, f >}
\newcommand{\scontexteqcq}{/<s, q = \rho, f >}
\newcommand{\symbcontexteprob}{\boxplus}
\newcommand{\bigsymbcontexteprob}{\mathop\boxplus}
\newcommand{\contexteprob}[5]{\bigsymbcontexteprob_{#1}\!\!<#2, #3 = #4, #5 >}
\newcommand{\contexteprobqcq}{\bigsymbcontexteprob_{p_i}\!\!<s_i, q_i = \rho_i, f_i>}

\newcommand{\spcontexteprob}[2]{/\bigsymbcontexteprob_{#1}#2}
\newcommand{\spcontexteprobbin}[3]{/ #2\symbcontexteprob_{#1}#3}
\newcommand{\contextestable}{\downarrow}
\newcommand{\trace}[1]{Tr(#1)}
\newcommand{\traceout}[3]{Tr_{#1/#2}(#3)}
\newcommand{\varpile}[1]{\mbox{Var$(#1)$}} 
\newcommand{\petitevarpile}[1]{\mbox{\scriptsize Var$(#1)$}} 
\newcommand{\pileconcat}{|} 
\newcommand{\pileajout}{.}
\newcommand{\pileajoutp}[2]{#2.#1}
\newcommand{\rmpile}[2]{#1\backslash#2}
\newcommand{\dom}[1]{\mbox{dom($#1$)}}
\newcommand{\surcharge}[2]{#1 \cup #2}
\newcommand{\domsoust}[2]{#1\backslash #2}
\newcommand{\codomsoustq}[2]{#1 \backslash #2}
\newcommand{\tailleseq}[1]{\mbox{size}(#1)}
\newcommand{\transfu}{\mathcal U}
\newcommand{\obs}{\mathcal O}
\newcommand{\mstdp}[2]{M_{std, #1}[#2]}
\newcommand{\mstd}[1]{M_{std, #1}}
\newcommand{\mplusmoins}{M_{+-}}

\newcommand{\adjoint}[1]{#1^\dagger}
\newcommand{\appli}[1]{\mathcal{T}_{#1}}
\newcommand{\applip}[2]{\appli{#1}(#2)}
\newcommand{\gate}[1]{\mbox{\it #1}}
\newcommand{\proc}[1]{\mbox{\bf #1}}
\newcommand{\sautdeligne}{\vspace{10pt}}
\newcommand{\tab}{\hspace{20pt}}
\newcommand{\n}{I\!\! N}
\newcommand{\ensemble}[1]{\{#1\}}
\newcommand{\card}[1]{|#1|}
\newcommand{\norme}[1]{\|#1\|}
\newcommand{\abs}[1]{|#1|}

\title{ A Probabilistic Branching Bisimulation for Quantum Processes }
\author{Marie {\sc Lalire}\footnote{Marie.Lalire@imag.fr} \\
Leibniz Laboratory\\
46, avenue F\'elix Viallet - 
38000 Grenoble, France}

\begin{document}
\maketitle

\begin{abstract}
Full formal descriptions of algorithms making use of quan\-tum principles must take into account both quan\-tum and classical computing components and assemble them so that they communicate and cooperate.
Moreover, to model concurrent and distributed quan\-tum computations, as well as quan\-tum communication protocols, quan\-tum to quan\-tum communications which move qubits physically from one place to another must also be taken into account.

Inspired by classical process algebras, which provide a framework for modeling cooperating
computations, a process algebraic notation is defined, which provides a homogeneous style to formal descriptions of concurrent and distributed computations comprising both quan\-tum and classical parts.
Based upon an operational semantics which makes sure that quantum objects, operations and 
communications operate according to the postulates of quantum mechanics, a probabilistic branching bisimulation is defined among processes considered as having the same behavior.
\end{abstract}

\section{ Introduction }
\label{sectionIntroduction}

Quantum algorithms are frequently described by means of quantum gate networks.
This has several drawbacks, for instance, gate networks do not allow descriptions of loops nor conditional executions of parts of networks. 
So as to overcome these difficulties, a few quantum programming languages have been developed, such as:
QCL \cite{Omer00QCL}, an imperative language designed by Bernhard \"Omer which aims at simulating quantum programs,
qGCL \cite{Zuliani01These} by Paolo Zuliani which allows the construction of proved correct quantum programs through a refinement method,
QPL \cite{Selinger03QPL}, a functional language designed by Peter Selinger with a denotational semantics, and a few others.
Several quantum $\lambda$-calculus have also been developed: for example \cite{Tonder03LambdaCalculus} by Andr\'e Van Tonder, which is based on a simplified linear $\lambda$-calculus and \cite{ArrighiDowek04Turku} by Pablo Arrighi and Gilles Dowek, which is a "linear-algebraic $\lambda$-calculus".
  
Cooperation between quantum and classical computations is inherent in quan\-tum algorithmics.
Teleportation of a qubit state from Alice to Bob \cite{BennettBrassard93Teleportation} is a good example of this cooperation. Indeed, Alice carries out a measurement, the result of which (two bits) is sent to Bob, and Bob uses this classical result to determine which quantum transformation he must apply.
Moreover, initial preparation of quantum states and measurement of quantum results are two essential forms of interactions between the classical and quantum parts of computations which a language must be able to express.
Process algebras are a good candidate for such a language since they provide a framework for modeling cooperating computations.
In addition, they have well defined semantics and permit the transformation of programs as well as the formal study and analysis of their properties.
A key point in their semantics is the definition of an equivalence relation on processes. Bisimulation is an adequate equivalence relation to deal with communicating processes since it relates processes that can execute the same flows of actions while having the same branching structure.

Simon Gay and Rajagopal Nagarajan have also developed CQP, a language to describe communicating quantum processes \cite{NagGay04Turku}. This language is based on $\pi$-calculus. An important point in their work is the definition of a type system, and the proof that the operational semantics preserves typing.

This paper presents first the main points of the definition of a Quantum Process Algebra (QPAlg).
Then, examples of short quantum programs are given in section \ref{sectionExamples}.
Finally, a bisimulation among processes is defined in section \ref{sectionBisimul}.


\section{ Quantum communicating processes }
\label{sectionQuantProcAlg}

The process algebra developed here is based upon process algebras such as CCS \cite{Milner89CommConc} and Lotos \cite{Bolognesi87IsoLotos}.
In this process algebra, basic actions are communications among processes (emission, denoted
 $g \envoi x$ and reception, denoted $g \recep x$, where $g$ is a communication gate) and quantum actions (unitary transformations and measurements).
To create a process from basic actions, the prefix operator "$\prefix$" is used: if $\alpha$ is an action and $P$, a process, $\alpha\prefix P$ is a new process which performs $\alpha$ first, then behaves as $P$.

There are two predefined processes: $\stopproc$, the process that cannot perform any transition, and $\term$, which performs a "$\delta$-transition" for signaling successful termination, and becomes $\stopproc$ ("$\delta$-transitions" are necessary in the semantics of sequential composition of processes).

The operators of the process algebra are: sequential composition ($P\seq Q$), parallel composition ($P\para Q$), conditional choice
($\debutbloc \cond{c_1}{P_1},\ldots,\cond{c_n}{P_n}\finbloc$)
and restriction ($P\restrict L$). As for sequential composition, process $Q$ is executed if process $P$ terminates successfully, that is to say if $P$ performs a $\delta$-transition.
The process $\debutbloc \cond{c_1}{P_1},\ldots,\cond{c_n}{P_n}\finbloc$, where $c_i$ is a condition and $P_i$ a process, evolves as a process chosen nondeterministically among the processes $P_j$ such that $c_j$ is true.
Restriction is useful for disallowing the use of some gates (the gates listed in $L$),
thus forcing internal communication within process $P$. 
Communication can occur between two parallel processes whenever a value emission in one of them and a value reception in the other one use the same gate name.

The main points of QPAlg involving the quantum world are developed in the rest of this section. The precise syntax and the main inference rules of the semantics are given in appendix \ref{annexQPA}. For more details, see \cite{LalireJorrand04Turku}.

\subsection{ Quantum variables }
 \label{subsecQuantVar}

For the purpose of this paper, we consider that there are two types of variables, one classical: {\it $\nattype$}, for variables taking integer values, and one quan\-tum: {\it $\qubittype$} for variables standing for qubits. An extended version of the process algebra would of course also include quantum registers and other types of variables.

In classical process algebras, variables are instantiated when communications between processes  occur and cannot be modified after their instantiation. As a consequence, it is not necessary to store their values. In fact, when a variable is instantiated, all its occurrences are replaced by the value received.

Here, quan\-tum variables stand for physical qubits. Applying a unitary transformation to a variable which represents a qubit modifies the state of that qubit.
This means that values of variables are modified. For that reason, it is necessary to keep track of both variable names and variable states.

Variables are declared, the syntax is:
$\debutbloc \declvar{x_1}{t_1},\ldots,\declvar{x_n}{t_n} \findeclvar P \finbloc$
 where  $x_1,\ldots,x_n$ is a list of variables, $t_1,\ldots,t_n$ are their types, and $P$ is a process which can make use of these classical and quan\-tum variables.
To simplify the rest of this paper, the names of variables will always be considered distinct. 

In the inference rules which describe the semantics of processes, the states of processes are process terms $P$ together with contexts $C$, of the form $P\spcontexte C$.
The main purpose of a context is to maintain the quan\-tum state,
stored as $q = \rho$ where $q$ is a sequence of quan\-tum variable names
and $\rho$ a density matrix representing their current quan\-tum state. 
Moreover, in order to treat classical variables in a similar way, 
modifications of classical variables are also allowed. So, for the same reason as in the case of quan\-tum variables, classical values are stored in the context.
Storing and retrieving classical values is represented by functions
$f: \mbox{\it names} \rightarrow\mbox{\it values}$.
The context must also keep track of the embedding of variable scopes.
To keep track of parallel composition,
this is done via a "cactus stack" structure of sets of variables, called the environment stack ($s$), which stores variable scopes and types. The set of all the variables in $s$ is denoted $\varpile s$,
"$\pileajout$" adds an element on top of a stack, and "$\pileconcat$" concatenates two stacks.

In summary, the context has three components $\contexteqcq$, where:
\begin{itemize}
\item $s$ is the environment stack;
\item $q$ is a sequence of quan\-tum variable names;
\item $\rho$ is a density matrix representing the quan\-tum state of the variables in $q$;
\item $f$ is the function which associates values with classical variables.
\end{itemize}

\sautdeligne
The rules for declaration and liberation of variables are the following:

\sautdeligne
\noindent
{\bf Declaration:}
$$
\frac{}{\debutbloc \declvar{x_1}{t_1},\ldots,\declvar{x_n}{t_n}\findeclvar
P \finbloc \spcontexte{C} \tautransition \debutbloc P \finbloc \spcontexte{C'}}
$$
with $C = \contexteqcq$, $C' = \contexte{s'}{q}{\rho}{f}$\\
and $s' = \pileajoutp{s} {\ensemble{(x_1,t_1),\ldots,(x_n,t_n)}}$

\sautdeligne
\noindent
This rule adds the new variable names and types on top of the stack $s$. Because the variables do not have values yet, the quan\-tum state and the classical function do not have to be modified at this point.

\sautdeligne
\noindent
{\bf Evolution of a process within the scope of declared variables:}
$$
\frac{P \spcontexte{C} \atransition{\alpha} P' \spcontexte{C'}}
{\debutbloc P \finbloc \spcontexte{C} \atransition{\alpha} \debutbloc P'\finbloc \spcontexte{C'}}
\hspace{10pt} \alpha \neq \delta
$$

In short:
if the process $P$ can perform a transition, then the process $\debutbloc P \finbloc$
can perform the same transition, provided that the action of the transition is not $\delta$. 

\sautdeligne
\noindent
{\bf Termination of a process with exit from a scope and liberation of the variables:}
$$
\frac{P \spcontexte{C} \dtransition P' \scontexte{\pileajoutp s e}{q}{\rho}{f}}
{\debutbloc P \finbloc \spcontexte{C} \dtransition \stopproc\ 
\scontexte{s}{\codomsoustq q {\varpile e}}
{\traceout{\petitevarpile e}{q}{\rho}}{\domsoust{f}{\varpile e}}}
$$

If the action is $\delta$, this means that $P$ has successfully terminated, so the context must be cleaned up by eliminating the variables having their scope limited to that process. 
These variables have their names listed in the head $e$ of the stack.
So, cleaning up the context means eliminating the head of the stack, removing the variables in $e$ from the sequence $q$ and from the domain of the function $f$.
The new quantum state is obtained by performing a partial trace on $\rho$ over the qubits in
$\varpile e$, which is denoted $\traceout{\petitevarpile e}{q}{\rho}$.

\subsection{ Basic actions }

The classical  basic actions are classical to classical communications and will not be further defined here. Classical to quan\-tum communications and quan\-tum to quan\-tum communications are introduced for respectively initializing qubits and allowing the description of quantum communication protocols. Quan\-tum to classical communications are part of measurement and are dealt with in the next paragraph.

The semantics of communications involving the quantum world is based upon the following rules
concerning the quantum side of such communications:
$$
\frac{}{g\envoi x\prefix P \spcontexte C \atransition{g\envoi x} P \spcontexte{C'}}
$$
where
\begin{itemize}
\item $C = \contexteqcq$ and $C' = \contexte{\rmpile s {\ensemble x}}
{\codomsoustq{q}{\ensemble x}}{\traceout{\ensemble x}{q}{\rho}}{f}$
\item $x\in\varpile s$ and $x\in q$
\end{itemize}
and:
$$
\frac{}{g \recep x\prefix P \spcontexte C \atransition{g\recep x} P\spcontexte C'}
$$
where
\begin{itemize}
\item $C=\contexteqcq$, $C'=\contexte{s}{x.q}{\nu\otimes\rho}{f}$
\item $x \in \varpile s$, $x$ of type $\qubittype$, $x \not\in q$
\item $\nu$ density matrix of dimension $2$
\end{itemize}

\sautdeligne
The first rule deals with qubit sending, and the other one, with reception of a qubit.
For qubit sending, because of the no-cloning theorem, the sent qubit must be removed from the context. The two rules concerning classical value sending and classical value reception are given in appendix \ref{annexQPASem}.

In the operational semantics of parallel composition, the combination of the rules for emission and reception defines communication.
In a classical to quan\-tum communication, the qubit is initialized in the basis state $\ket v \bra v$, where $v$ is the classical value sent (in this case, $v$ must be $0$ or $1$). In a quantum to quantum communication, the name of the sent qubit is replaced in $q$ by the name of the receiving qubit.

\sautdeligne
Other basic actions are unitary transformations which perform the unitary evolution of qubit states. Given a set $\transfu$ of predefined unitary transformations, the action corresponding to the application of $U\in\transfu$ to a list of quan\-tum variables is denoted by $U[x_1,\ldots,x_n]$. 

The inference rule for unitary transformations is:
$$
\frac{}{U[x_1,\ldots,x_n]\prefix P \scontexteqcq \tautransition
P \scontexte{s}{q}{\rho'}{f}}
$$
where
\begin{itemize}
\item  $U\in \transfu$, $x_1,\ldots,x_n \in \varpile s$, and $x_1,\ldots,x_n \in q$
\item $x_1,\ldots,x_n$ are pairwise distinct
\item $\rho' = \applip{U}{\rho}$
\end{itemize}

The condition $x_1,\ldots,x_n \in q$ prevents from applying a unitary transformation to qubits which have not been initialized.

$\appli{U}$ is the super-operator which must be applied to $\rho$, to describe the evolution of the quantum state due to the application of the unitary transformation $U$ to the qubits $x_1,\ldots,x_n$. In general, with $A$ a $2^n\times 2^n$ matrix:
$$\appli{A}: \rho \mapsto\adjoint{\Pi}.(A\otimes I^{\otimes k}).\Pi.\rho.\adjoint{\Pi}.
(\adjoint{A}\otimes I^{\otimes k}).\Pi$$
where
\begin{itemize}
\item $\Pi$ is the permutation matrix which places the $x_i$'s at the head of $q$
\item $k = \tailleseq q - n\ $
\item $I^{\otimes k}\!= \underbrace{I \otimes \cdots \otimes I}_k$, where $I$ is the identity matrix on $\mathbb C^2$
\end{itemize}

Since the unitary transformation $U$ may be applied to qubits which are anywhere within the list $q$, a permutation $\Pi$ must be applied first. This permutation moves the $x_i$'s so that they are placed at the head of $q$ in the order specified by $[x_1,\ldots,x_n]$.
Then $U$ can be applied to the first $n$ elements and $I$ to the remainder. Finally, the last operation is the inverse of the permutation $\Pi$
so that at the end, the arrangement of the elements in $\rho$ is consistent with the order of the elements in $q$.

\subsection{ Measurement and probabilistic processes }

Last but not least, an essential basic action has to be introduced into the process algebra: quan\-tum measurement.
Let $M$ be an observable in a set $\obs$ of predefined observables, $x_1,\ldots,x_n$ a list of distinct quan\-tum variables and $g$ a gate. The syntax for measurement is the following:
\begin{itemize}
\item $M[x_1,\ldots,x_n]$ is a measurement of the $n$ qubits of the list with respect to observable $M$, but the classical result is neither stored nor transmitted.
\item $g \envoi{M[x_1,\ldots,x_n]}$ is a measurement of the $n$ qubits of the list with respect to observable $M$, followed by sending the classical result through gate $g$.
\end{itemize}

Measurement is probabilistic: more precisely, the classical result and the quan\-tum state after measurement are probabilistic.
In the case of measurement without communication of the classical result, only the quantum state is probabilistic after measurement, so the probabilities can be reflected in the density matrix:
$$
\frac{}{M[x_1,\ldots,x_n]\prefix P \scontexteqcq \tautransition
P\scontexte{s}{q}{\rho'}{f}}
$$
with
\begin{itemize}
\item $x_1,\ldots,x_n \in \varpile s$, $x_1,\ldots,x_n \in q$ and
$x_1,\ldots,x_n$ are pairwise distinct
\item $M\in\obs$  with $\sum_{i} \lambda_i P_i$ as spectral decomposition
\item $\rho' =  \sum_{i}\applip{P_i}{\rho}$
\end{itemize}
As in the case of unitary transformations, $\appli{P_i}$ is the super-operator corresponding to the application of the projector $P_i$ to measured qubits.
The computation of $\rho'$ stems from the projective measurement postulate of quan\-tum mechanics.

When the value coming out of the measurement is sent out, the classical result is probabilistic.
This requires the introduction of a probabilistic composition operator for contexts.
This operator is denoted $\symbcontexteprob_p$:
the state $P\spcontexteprobbin{p}{C_1}{C_2}$ is $P\spcontexte{C_1}$ with probability $p$ and $P\spcontexte{C_2}$ with probability $1-p$.
In general, a context is either of the form $\contexteqcq$, or of the form
$\contexteprobqcq$ where the $p_i$'s are probabilities adding to $1$.
Then, the rule for measurement followed by sending the classical result is:
$$
\frac{}{g\envoi{M[x_1,\ldots,x_n]} \prefix P \spcontexte C \tautransition
\debutbloc g\envoi y\prefix \term \finbloc \seq P\spcontexteprob{p_i}{C_i}}
$$
where
\begin{itemize}
\item $C = \contexteqcq$
\item $C_i = \contexte{\pileajoutp s {\ensemble{(y,\nattype)}}}{q}{\rho_i}
{\surcharge f \ensemble{y\mapsto\lambda_i}}$
\item $x_1,\ldots,x_n \in \varpile s$, $x_1,\ldots,x_n \in q$ and
$x_1,\ldots,x_n$ are pairwise distinct
\item $y$ is a new variable (introduced as $\cvar y$ by this rule)
\item $M\in\obs$  with $\sum_{i} \lambda_i P_i$ as spectral decomposition
\item $p_i = \trace{\applip{P_i}{\rho}}$, \hspace{10pt} $\rho_i = \frac{1}{p_i}\applip{P_i}{\rho}$
\end{itemize}

As explained in \cite{Cazorla01Art} and \cite{CazorlaMis}, if a process contains both a probabilistic and a nondeterministic choice, the probabilistic choice must always be solved first.
In the process algebra presented here, nondeterminism appears with parallel composition and conditional choice. So as to guarantee that probabilistic choice is always solved first, the notion of probabilistic stability for contexts is introduced: a context $C$ is probabilistically stable, which is denoted $C\contextestable$, if it is of the form $\contexteqcq$.
If the context of a process state is not stable, a probabilistic transition must be performed first:
$$
\frac{}{P\spcontexteprob{p_i}{C_i} \ptransition{p_j} P\spcontexte{C_j}}
\mbox{ where } \sum_i p_i = 1
$$
where $S_1 \ptransitionse p S_2$ means that state $S_1$ becomes $S_2$ with probability $p$.


\section{ Examples }
\label{sectionExamples}

\newcommand{\eprinit}{\proc{BuildEPR}}
\newcommand{\verifepr}{\proc{CheckEPR}}
\newcommand{\alice}{\proc{Alice}}
\newcommand{\bob}{\proc{Bob}}
\newcommand{\teleport}{\proc{Teleport}}
\newcommand{\meas}{\gate{meas}}
\newcommand{\eve}{\proc{Eve}}
\newcommand{\canal}{\proc{Channel}}
\newcommand{\faille}{\proc{Flaw}}
\newcommand{\prot}{\proc{Protocol}}
\newcommand{\aaa}{\proc{A}}
\newcommand{\bbb}{\proc{B}}
\newcommand{\eee}{\proc{E}}
\newcommand{\fillcanal}{\gate{fill}}
\newcommand{\emptycanal}{\gate{empty}}
\newcommand{\fillfaille}{\gate{fillFlaw}}
\newcommand{\emptyfaille}{\gate{emptyFlaw}}
\newcommand{\random}{\proc{Random}}
\newcommand{\aaaun}{\proc{A$_1$}}
\newcommand{\aaadeux}{\proc{A$_2$}}
\newcommand{\dataa}{\gate{dataA}}
\newcommand{\datab}{\gate{dataB}}
\newcommand{\basea}{\gate{baseA}}
\newcommand{\baseb}{\gate{baseB}}
\newcommand{\bool}{\gate{bool}}
\newcommand{\ok}{\gate{ok}}
\newcommand{\keepdataa}{\gate{keepDataA}}
\newcommand{\keepdatab}{\gate{keepDataB}}
\newcommand{\base}{\gate{base}}
\newcommand{\keep}{\gate{keep}}
\newcommand{\received}{\gate{received}}

In the following examples, the set $\transfu$ of unitary transformations is:
$$\transfu = \{H, CNot, I, X, Y, Z\}$$
where $H$ is Hadamard transformation, $CNot$ is "controlled not", $I$ is the identity, and $X, Y, Z$ are Pauli operators.
The set $\obs$ of observables contains the observables corresponding to measurement of one and two qubits in the standard basis, denoted respectively $\mstd{1}$ and $\mstd{2}$, and the observable corresponding to measurement of a qubit in the basis $\ensemble{\ket{+},\ket{-}}$, denoted $\mplusmoins$.

\subsection{Teleportation}

Once upon a time, there were two friends, Alice and Bob who had to separate and live away from each other.
Before leaving, each one took a qubit of the same EPR pair.
Then Bob went very far away, to a place that Alice did not know.
Later on, someone gave Alice a mysterious qubit in a state $\ket\psi = \alpha\ket 0+\beta\ket 1$,
with a mission to forward this state to Bob.
Alice could neither meet Bob and give him the qubit, nor clone it and broadcast copies everywhere, nor obtain information about $\alpha$ and $\beta$. Nevertheless, Alice succeeded thanks to the EPR pair and the teleportation protocol \cite{BennettBrassard93Teleportation}:

\sautdeligne
$\begin{array}{lcl}
\eprinit & \procdef & \debutbloc \qvar x, \qvar y \findeclvar \\
&& \tab (( g_1\recep{x} \prefix g_2 \recep y \prefix H[x] \prefix CNot[x,y] \prefix\term)\\
&& \tab \para (g_1\envoi 0\prefix g_2\envoi 0\prefix\term)) \restrictg{g_1,g_2}\\
&& \finblocse \\
\end{array}$

\sautdeligne
$\begin{array}{lcl}
\alice&\procdef& \debutbloc \qvar x, \qvar y \findeclvar\\
&& \tab CNot[x,y] \prefix H[x] \prefix \meas \envoi{\mstdp{2}{x,y}} \prefix \term\\
&& \finblocse \\
\end{array}$

\sautdeligne
$\begin{array}{lcl}
\bob&\procdef&\debutbloc \qvar z \findeclvar\\
&& \tab \debutbloc \cvar k \findeclvar\\
&& \tab\tab \meas\recep k\prefix\\
&& \tab\tab \debutbloc \cond{k=0}{I[z]\prefix\term}, \\
&& \tab\tab \cond{k=1}{X[z]\prefix\term}, \\
&& \tab\tab \cond{k=2}{Z[z]\prefix\term}, \\
&& \tab\tab \cond{k=3}{Y[z]\prefix\term} \finbloc \\
&&\tab \finblocse\\
&& \finblocse \\
\end{array}$

\sautdeligne
$\begin{array}{lcl}
\teleport&\procdef& \debutbloc \qvar\psi \findeclvar\\
&& \tab \debutbloc \qvar a, \qvar b \findeclvar\\
&& \tab\tab \eprinit[a,b]\seq\\
&& \tab\tab (\alice[\psi,a]\para\bob[b])\restrictg\meas\\
&& \tab \finblocse\\
&& \finblocse\\
\end{array}$

\sautdeligne
The inference rules can be used to show that this protocol results in Bob's $z$ qubit having the state initially possessed by the $x$ qubit of Alice, with only two classical bits sent from Alice to Bob.

\subsection{ Communication protocols }

Alice sends qubits to Bob through a non secure channel and Eve eavesdrops this channel to get information on the qubits sent by Alice.
In the following example $\aaa$, $\bbb$, and $\eee$ are processes modeling whatever Alice, Bob, and Eve may respectively apply to their qubits. The actions of these processes, which are not made explicit here, will be specified in the next example of the BB84 protocol.

The communication protocols which are described here could be used to mo\-del cryptographic protocols so as to check if they are secure.

\sautdeligne
\subsubsection*{Eve intercepts all qubits}

Eve intercepts qubits because of a flaw in the channel that Alice and Bob are using to communicate.

\sautdeligne
$\begin{array}{lcl}
\alice & \procdef & \debutbloc \qvar a \findeclvar
\aaa[a]\seq \fillcanal \envoi a \prefix\term \finbloc \seq \alice\\
\end{array}$

\sautdeligne
$\begin{array}{lcl}
\bob &\procdef& \debutbloc \qvar b \findeclvar
\emptycanal \recep b \prefix \bbb[b] \finbloc \seq \bob\\
\end{array}$

\sautdeligne
$\begin{array}{lcl}
\eve &\procdef& \debutbloc \qvar e, \qvar f \findeclvar\\
&& \tab \emptyfaille \recep e \prefix \eee[e,f] \seq \fillfaille \envoi f \prefix \term\\
&& \finblocse \seq \eve\\
\end{array}$

\sautdeligne
$\begin{array}{lcl}
\faille &\procdef& \debutbloc \qvar u, \qvar v\findeclvar
\emptyfaille\envoi u \prefix \fillfaille \recep v \prefix\term \finbloc\\
\end{array}$

\sautdeligne
$\begin{array}{lcl}
\canal &\procdef& \debutbloc \qvar x, \qvar y\findeclvar
\fillcanal \recep x \prefix\faille[x,y] \seq \emptycanal\envoi y \prefix\term \finbloc \seq\\
&&\canal\\
\end{array}$

\sautdeligne
$\begin{array}{lcl}
\prot &\procdef& (\alice\para\bob\para\eve\para\canal)\\
&& \tab\restrict \{\fillcanal,\emptycanal,\fillfaille,\emptyfaille\}
\end{array}$

\sautdeligne
\subsubsection*{Eve intercepts some of the qubits}

This part assumes that a nondeterministic process composition $P\nondet Q$ is introduced in the process algebra. This operator is not presented in the operational semantics in appendix \ref{annexQPASem} but it can be simulated by
$\debutbloc \cond{\mbox{\it true}}{P},\cond{\mbox{\it true}}{Q}\finbloc$.

To model the fact that Eve does not succeed in intercepting all qubits, the flaw in the channel is made nondeterministic:
$$
\begin{array}{lcl}
\canal &\procdef& \debutbloc \qvar x \findeclvar\\
&& \tab \fillcanal \recep x \prefix\\
&& \tab (\\
&& \tab\tab \debutbloc \qvar y \findeclvar \faille[x,y]\seq\emptycanal\envoi y\prefix\term \finbloc
\nondet\\
&& \tab\tab (\emptycanal\envoi x \prefix\term)\\
&& \tab )\\
&& \finblocse \seq \canal
\end{array}
$$

\subsection{The BB84 protocol}

The BB84 protocol \cite{BennettBrassard84BB84} is a protocol for secure quan\-tum key distribution: Alice and Bob must agree on a private key, i.e. a list of bits that should remain secret. To communicate, they send qubits through a non secure channel. In fact, the processes $\aaa$ and $\bbb$ left unspecified in the previous paragraph can be used to model this protocol. 
The process $\alice$ is redefined and the process $\bbb$ used by $\bob$ is made explicit.
In addition, another process is defined: the process $\random$ which initializes a bit randomly at $0$ or $1$.
The gates $\keepdataa$ and $\keepdatab$ are used by Alice and Bob respectively to send the bits that they want to keep.

\sautdeligne
$\begin{array}{lcl}
\alice & \procdef & \debutbloc \qvar{a}, \cvar{\dataa}, \cvar{\basea} \findeclvar\\
&&\tab \aaaun[a,\dataa,\basea] \seq \fillcanal \envoi a \prefix  \aaadeux[\dataa,\basea]\\
&&\finblocse \seq \alice
\end{array}$

\sautdeligne
$\begin{array}{lcl}
\random &\procdef& \debutbloc \cvar r \findeclvar\\
&& \tab \debutbloc \qvar x \findeclvar\\
&& \tab\tab (g\envoi 0\prefix\term \para g\recep x\prefix\term)\restrictg{g} \seq\\
&& \tab\tab H[x]\prefix\\
&& \tab\tab (h \envoi{\mstdp 1 x}\prefix\term \para h\recep r\prefix\term)\restrictg{h}\\
&& \tab \finblocse \\
&& \finblocse \\
\end{array}$

\sautdeligne
$\begin{array}{lcl}
\aaaun & \procdef & \debutbloc \qvar a, \cvar \dataa, \cvar \basea \findeclvar\\
&& \tab \random[\dataa][a]\seq\\
&& \tab \random[\basea]\seq\\
&& \tab \debutbloc \cond{\basea=1}{H[a]\prefix\term} \finbloc \\
&&  \finblocse \\
\end{array}$

\sautdeligne
$\begin{array}{lcl}
\aaadeux & \procdef & \debutbloc \cvar\dataa,\cvar\basea \findeclvar\\
&& \tab \debutbloc \cvar {\bool}, \cvar\ok \findeclvar \\
&& \tab\tab \received \recep \ok \prefix\\
&& \tab\tab \base \envoi \basea \prefix\\
&& \tab\tab \keep \recep \bool \prefix\\
&& \tab\tab \debutbloc \cond{\bool=1}{\keepdataa \envoi \dataa \prefix\term} \finbloc \\
&& \tab \finblocse \\
&& \finblocse \\
\end{array}$

\sautdeligne
$\begin{array}{lcl}
\bbb &\procdef& \debutbloc \qvar b \findeclvar\\
&& \tab \debutbloc \cvar\basea, \cvar\baseb, \cvar\datab \findeclvar\\
&& \tab\tab \random[\baseb]\seq\\
&& \tab\tab (\\
&& \tab\tab\tab\debutbloc \cond{\baseb=0}{g\envoi{\mstdp 1 b}\prefix\term},\\
&& \tab\tab\tab\ \ \cond{\baseb=1}{g\envoi{M_{+-}[b]}\prefix\term} \finbloc \\
&& \tab\tab\tab \para g \recep \datab \prefix \term\\
&& \tab\tab )\restrictg g\seq\\

&& \tab\tab \received \envoi 1 \prefix\\
&& \tab\tab \base \recep \basea \prefix\\
&& \tab\tab \debutbloc
\cond{\basea=\baseb}{\keep \envoi 1\prefix\keepdatab \envoi{\datab}\prefix \term},\\
&& \tab\tab \ \ \cond{\basea\neq\baseb}{\keep \envoi 0 \prefix \term} \finbloc\\
&& \tab \finblocse\\
&& \finblocse\\
\end{array}$


\section{ Probabilistic branching bisimulation }
\label{sectionBisimul}

\newcommand{\bisimul}{\equiv} 
\newcommand{\ensetats}{\EuScript{S}}
\newcommand{\siltransition}{\leadsto}
\newcommand{\seqsiltransition}{\siltransition^*}
\newcommand{\equivpbb}{\leftrightarroweq_{pb}}
\newcommand{\etatqvar}[2]{\rho_{#1}^{#2}}
\newcommand{\classeeq}[1]{\bar{#1}}
\newcommand{\parties}[1]{\mathcal{P}(#1)}
\newcommand{\atteind}{\triangleright}

The operational semantics associates a process graph with a process state. A process graph is a graph where vertices are process states and edges are transitions labeled with actions or probabilities. Each process graph has an initial state.

A bisimulation is an equivalence relation on process states. It identifies states when they are associated with process graphs having the same branching structure.

The bisimulation defined here is probabilistic because of probabilities introduced by quantum measurement and branching because some transitions are considered as silent.
It is inspired from the definitions in \cite{Fokkink00PA} and \cite{Andova99Art}.

\subsection{ Preliminary definitions and notations }

\subsubsection*{ Process states }

The set of all possible process states is denoted $\ensetats$.
Let $S, T \in \ensetats$, then $S$ can be written $P\spcontexte{C_P}$ and  $T$, $Q\spcontexte{C_Q}$ where $P$, $Q$ are process terms and $C_P$, $C_Q$ contexts (possibly probabilistic).

Assuming that $S=P\spcontexte{C_P}$ and $C_P=\contexteqcq$,  if $x$ is a qubit in $S$ and $x\in q$, then $\etatqvar x S$ is the state of $x$ and this state can be obtained with a trace out operation on $\rho$:
$$\etatqvar x S = \traceout{\ensemble x}{q}{\rho}$$ 
 
\subsubsection*{ Silent transitions }

The transitions considered as silent are of course internal transitions ($\petitetautransition$) but also probabilistic transitions. The reason is that we want, for example, the following states $S_1$ and $S_2$ to be equivalent.

\begin{center}
\begin{tabular}{c}
\begin{picture}(100,50)
\put(50,35){\vector(0,-1){20}}
\put(53,25){\mbox{$a$}}
\put(46,40){\mbox{$S_1$}}
\put(47,5){\mbox{$T$}}
\end{picture}
\end{tabular}
\begin{tabular}{c}
\begin{picture}(100,75)
\put(46,65){\mbox{$S_2$}}
\put(50,60){\vector(1,-1){25}}
\put(50,60){\vector(-1,-1){25}}
\put(16,45){\mbox{\footnotesize 0.2}}
\put(76,45){\mbox{\footnotesize0.8}}
\put(25,35){\vector(0,-1){20}}
\put(75,35){\vector(0,-1){20}}
\put(28,25){\mbox{$a$}}
\put(78,25){\mbox{$a$}}
\put(22,5){\mbox{$T$}}
\put(72,5){\mbox{$T$}}
\end{picture}
\end{tabular}
\end{center}

Silent transitions will be denoted $\siltransition$. $\seqsiltransition$ stands for a sequence (possibly empty) of silent transitions.

\subsubsection*{ Function $\mu$ }

Probabilistic transitions are considered silent, nonetheless, in two equivalent sta\-tes, the corresponding actions that can be performed on both sides must occur with the same probability.

Let $\equiv$ be an equivalence on process states,
$S$ be a process state and $\classeeq S$, its equivalence class with respect to $\equiv$.
If $M$ is a set of process states and $S$ a state, then $S \atteind M$ means that there exists a
sequence of transitions remaining in $M \cup \classeeq{S}$, from $S$ to a state of $M$.

A function $\mu_{\equiv} : \ensetats \times \parties{\ensetats} \rightarrow [0,1]$ is defined for computing the probability to reach a state in the set $M$ from a state $S$ without leaving $\classeeq{S} \cup M$.
It should be noted that, for this function to yield a probability, nondeterminism must be eliminated in a way which allows the computation of $\mu$.
Here, nondeterminism is treated as equiprobability, but this is just a convention for the definition of  $\mu_{\equiv}$.
For example, it does not imply the equivalence of the following two process states:
\begin{center}
\begin{tabular}{c}
\begin{picture}(100,50)
\put(50,35){\vector(1,-1){22}}
\put(50,35){\vector(-1,-1){22}}
\put(30,25){\mbox{$a$}}
\put(65,25){\mbox{$a$}}
\put(46,40){\mbox{$S_1$}}
\put(22,5){\mbox{$T$}}
\put(72,5){\mbox{$T$}}
\end{picture}
\end{tabular}
\begin{tabular}{c}
\begin{picture}(100,75)
\put(46,65){\mbox{$S_2$}}
\put(50,60){\vector(1,-1){25}}
\put(50,60){\vector(-1,-1){25}}
\put(16,45){\mbox{\footnotesize 0.5}}
\put(76,45){\mbox{\footnotesize 0.5}}
\put(25,35){\vector(0,-1){20}}
\put(75,35){\vector(0,-1){20}}
\put(28,25){\mbox{$a$}}
\put(78,25){\mbox{$a$}}
\put(22,5){\mbox{$T$}}
\put(72,5){\mbox{$T$}}
\end{picture}
\end{tabular}
\end{center}

A bisimulation is an equivalence relation $\equiv$ which must verify some properties, among which: if $S \equiv T$ then $\mu_{\equiv}(S,M) = \mu_{\equiv}(T,M)$ for all $M$ equivalence class of  $\equiv$.
In this case, $\mu_{\equiv}(S,M)$ is the probability to perform an action.

The function $\mu_{\equiv}$ is defined by:

\sautdeligne
\noindent
\begin{tabular}{lr}
\begin{tabular}{l}
$\bullet$\hspace{5pt} if $S \in M$ then $\mu_{\equiv}(S,M) = 1$ 
\end{tabular}&\\
\\ %
\begin{tabularx}{0.62\linewidth}{X}
$\bullet$\hspace{5pt} 
else if $\exists\ T \in M \cup \classeeq S$ such that
$S \ptransitionse p T \atteind M$ then
let $E_S = \ensemble{R\in M\cup\classeeq{S}\ |\ S \ptransitionse{p_{_R}} R \atteind M}$ in
$$\mu_{\equiv}(S,M) = \sum_{R\in E_S} p_{_R}\  \mu_{\equiv}(R,M)$$
\end{tabularx}
&
\begin{tabular}{c}
\includegraphics[scale=0.7]{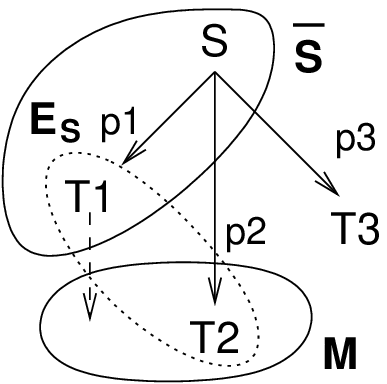}
\end{tabular}\\
\\ %
\begin{tabularx}{0.62\linewidth}{X}
$\bullet$\hspace{5pt}
else if $\exists\ T \in M \cup \classeeq S$ such that
$S \petiteatransition{a} T \atteind M$ then
let $E_S = \ensemble{R\in M\cup\classeeq{S}\ |\ S \petiteatransition{a_{_{R}}} R \atteind M}$ in
$$\mu_{\equiv}(S,M) = \frac{1}{\card{E_S}} \sum_{R\in E_S} \mu_{\equiv}(R,M)$$
\end{tabularx}
&
\begin{tabular}{c}
\includegraphics[scale=0.7]{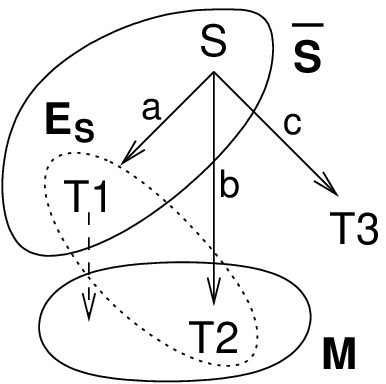}
\end{tabular}\\
\\ %
\begin{tabular}{l}
$\bullet$\hspace{5pt} else $\mu_{\equiv}(S,M) = 0$ 
\end{tabular}&\\
\end{tabular}

\subsubsection*{ Equivalence of contexts }

Let $\sigma$ be a renaming on variables. The extensions of $\sigma$ to environment stacks and sequences of quantum variables are also called $\sigma$.

Let $C=\contexteqcq$ and $C'=\contexte{s'}{q'}{\rho'}{f'}$ be two contexts,
$C$ and $C'$ are equivalent, if and only if there exists a renaming $\sigma$ such that:
\begin{itemize}
\item $\sigma(s) = s'$
\item $\exists$ a permutation $\pi$ such that $\pi(\sigma(q))=q'$ and $\pi(\rho)=\rho'$
\item $\forall x \in \dom{f}$, $\sigma(x) \in \dom{f'}$ and $f(x)=f'(\sigma(x))$
\item $\forall y \in \dom{f'}$, $\sigma^{-1}(y) \in \dom{f}$ and $f(\sigma^{-1}(y))=f'(y)$
\end{itemize}

This equivalence relation can easily be extended to probabilistic contexts.

\subsection{ Probabilistic branching bisimulation: definition }

An equivalence relation is a probabilistic branching bisimulation if and only if:
\renewcommand{\arraystretch}{1.5}

\vspace{5pt}
\noindent
\begin{tabular}{ll}
\begin{tabularx}{0.6\linewidth}{cX}
$\ \bullet$& if $S$ and $T$ are equivalent and if an action $a$ can be performed from $S$, then the same action can performed from $T$, possibly after several silent transitions;\\
$\ \bullet$& the reached states ($S'$ and $T''$) are equivalent;\\
$\ \bullet$& the action $a$ must occur with the same probability in both cases.
\end{tabularx}
&
\begin{tabular}{c}
\includegraphics[scale=0.7]{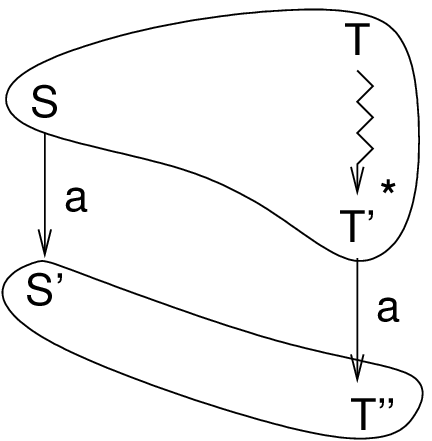}
\end{tabular}
\end{tabular}
\renewcommand{\arraystretch}{1}

\sautdeligne
Let $\bisimul$ be an equivalence relation on process states.
$\bisimul$ is a probabilistic branching bisimulation if and only if it satisfies:

\begin{itemize}
\item {\bf Termination}

if $S \bisimul T$ and $S \petitedtransition S'$ then\\
$\exists\  T', T''$ such that
$(T\seqsiltransition T' \petitedtransition T''
\ \wedge\  S\bisimul T' \ \wedge\  S'\bisimul T'')$ and the contexts in $S'$ and $T''$ are equivalent

\item {\bf Value sending}

if $S \bisimul T$ and $S \petiteatransition{g\envoi v} S'$ ($v$ classical value) then\\
$\exists\  T', T''$ such that
$(T\seqsiltransition T' \petiteatransition{g\envoi v} T''
\ \wedge\  S\bisimul T' \ \wedge\  S'\bisimul T'')$

\item {\bf Qubit sending}

if $S \bisimul T$ and $S \petiteatransition{g\envoi x} S'$ ($x$ variable) then\\
$\exists\  T', T''$ such that
$(T\seqsiltransition T' \petiteatransition{g\envoi y} T''
\ \wedge\  \etatqvar{x}{S}=\etatqvar{y}{T}
\ \wedge\  S\bisimul T' \ \wedge\  S'\bisimul T'')$

\item {\bf Value reception}

if $S \bisimul T$ and $S \petiteatransition{g\recep v} S'$ ($v$ classical value) then\\
$\exists\  T', T''$ such that
$(T\seqsiltransition T' \petiteatransition{g\recep v} T''
\ \wedge\  S\bisimul T' \ \wedge\  S'\bisimul T'')$

\item {\bf Qubit reception}

if $S \bisimul T$ and $S \petiteatransition{g\recep x} S'$ ($x$ variable) then\\
$\exists\  T', T''$ such that
$(T\seqsiltransition T' \petiteatransition{g\recep y} T''
\ \wedge\  \etatqvar{x}{S'}=\etatqvar{y}{T''}
\ \wedge\  S\bisimul T' \ \wedge\  S'\bisimul T'')$

\item {\bf Silent transition}

if $S \bisimul T$ and $S \siltransition S'$ then $S' \bisimul T$

\item {\bf Probabilities}

if $S \bisimul T$ then $\mu_{\bisimul}(S,M) = \mu_{\bisimul}(T,M),\ \forall M \in (\ensetats/\bisimul)\backslash\ensemble{\classeeq S}$
\end{itemize}

\subsection{ Bisimulation and recursion }

Recursion in a process definition introduces circuits in the associated process graph.
As a consequence, it must be proved that in this case, the notion of probabilistic branching bisimulation is well-defined. In fact, the only point that could be a problem is the definition of $\mu_{\equiv}$.

Let $S$ be a process state and $M$ be a set of process states, the computation of
$\mu_{\bisimul}(T,M)$ for all $T$ in $\classeeq{S}$ leads to a linear system of equations where the 
$\mu_{\bisimul}(T,M)$ are the unknowns:
$$X = AX + B$$

The $i^{th}$ row in this system can be written:
$x_i = a_{i1} x_1 + \cdots + a_{in} x_n + b_i$.

The coefficients $a_{ij}$ and $b_i$ are either probabilities or average coefficients in the case of nondeterminism. As a consequence: $0 \leq a_{ij} \leq 1$ and $0 \leq b_{i} \leq 1$.
Moreover, in the definition of $\mu_{\equiv}$, every state in the set $E_S$ is such that there exists a path from that state to the set $M$. 
Therefore, the system of equations obtained can be transformed into a system such that $b_i > 0$, $\forall i \in \llbracket 1,n \rrbracket$.
From now on, we consider that the system verifies this property.

Another property of the system is: $0 \leq \sum_j a_{ij} + b_i \leq 1$, thus $0 \leq \sum_j a_{ij} < 1$.

To prove that the system has a unique solution, it is sufficient to prove $\norme{A}<1$ and use the fixpoint theorem.
The norms for matrices and vectors are:
$$\norme{A} = \sup_{\norme{X}=1} \norme{AX} \hspace{20pt} \norme{X} = \max_{i} \abs{x_i}$$
We obtain:
$$\norme{AX} = \max_{i}\abs{\sum_{j} a_{ij} x_i} \leq \max_i \sum_j (\abs{a_{ij}}\abs{x_i})$$
$$\mbox{and then } \norme{A} \leq \max_i \sum_j \abs{a_{ij}} < 1$$

$\norme{A}<1$ implies that the function $f: X \mapsto AX+B$ is strictly contracting, so
from the fixpoint theorem, we infer that the equation $X =AX+B$ has a unique solution.
Moreover, as $f([0,1]) \subseteq [0,1]$, this solution belongs to $[0,1]$.

As a consequence, the function $\mu_{\equiv}$ is well-defined even in case of recursive processes.


\section{ Conclusion }

This paper has presented a process algebra for quantum programming which can describe both classical and quantum programming, and their cooperation.
Without this cooperation, the implementation of protocols like BB84 is not possible.
Another feature of this language is that measurement and initialization of quantum registers occur through communications between quantum and classical parts of the language, which happens to be a faithful model of physical reality.

Moreover, a thorough semantics has been defined, thus allowing the study and analysis of programs. One peculiarity of  this  semantics is the introduction of probabilistic processes, due to quantum measurement. Probabilistic processes perform probabilistic transitions. As a consequence, the execution tree obtained  from a process presents action branches and probabilistic branches.

Finally a semantical equivalence relation on processes has been defined. This equivalence is a bisimulation which identifies processes associated with process graphs having the same branching structure. This is the first step toward the verification of quantum cryptographic protocols.

Several extensions are possible. As already mentioned, a nondeterministic process composition operator can be introduced. A probabilistic composition of processes could be added. This would allow, for example, the description of communication protocols in which Eve intercepts qubits with a given probability.




\appendix

\section{The quantum process algebra}
\label{annexQPA}

\subsection{Syntax}
\label{annexQPASyntax}

\begin{tabular}{lcl}
\it process & $\syntaxdef$ & \it $\pmb\stopproc$\\
                  &$|$& \it $\pmb\term$ \\
                  &$|$& \it action $\pmb\prefix$ process\\
                  &$|$& \it  process $\pmb\seq$ process\\
                  &$|$& \it  process $\pmb\para$ process\\
                  &$|$& \it  process $\pmb \backslash \pmb\{$ gate\_list $\pmb\}$ \\
                  &$|$& \it  $\pmb\debutbloc$ cond\_list $\pmb \finbloc$\\
                  &$|$& \it $\pmb \debutbloc$ var\_decl\_list $\pmb \findeclvar$ process $\pmb\finbloc$ \\
                  &$|$& \it process\_name $[\pmb [$ var\_list $\pmb ]]$\\
&&\\
\it action & $\syntaxdef$ & \it communication \\
		&$|$&\it unit\_transf\\
		&$|$&\it measurement \\
&&\\
\it communication & $\syntaxdef$ & \it gate {\bf !} exp\\
                  &$|$&\it gate {\bf !} measurement\\
 		&$|$& \it gate {\bf ?} variable  \\
&&\\
\it unit\_transf         & $\syntaxdef$ & \it unitary\_operator $\pmb [$ var\_list $\pmb ]$ \\
&&\\
\it measurement    & $\syntaxdef$ & \it observable $\pmb [$ var\_list $\pmb ]$\\
&&\\
\it var\_decl & $\syntaxdef$ & \it variable $\pmb :$ var\_type\\
&&\\ 
\it proc\_def & $\syntaxdef$ & \it process\_name $\displaystyle\pmb{\procdef}$ process\\
\end{tabular}

\subsection{Main inference rules of the semantics}
\label{annexQPASem}

The semantics is specified with inference rules which give the evolution of the states of processes. There are four kinds of transitions:
\begin{itemize}
\item action transition: $\petiteatransition\alpha$ where $\alpha$ is $g\envoi x$ or $g\recep x$;
\item silent transition: $\petitetautransition$, for internal transition;
\item delta transition: $\petitedtransition$, for successful termination;
\item probabilistic transition: $\ptransitionse p$, where $p$ is a probability.
\end{itemize}

In the following, $P, Q, P', Q', P_i$ and $P_i'$ are processes, $C$, $C'$ and $C_i$ are contexts, $\alpha$ is an action, $g$ is a communication gate, $v$ is a value, $x$ is a variable, and $c_j$ is a condition.

\subsubsection*{Successful termination}
$$
\frac{}{\term\spcontexte C \dtransition \stopproc\spcontexte C}
\hspace{10pt}C\contextestable
$$
\subsubsection*{Action Prefix}
$$
\frac{}{g \envoi v \prefix P \spcontexte C \atransition{g\envoi v} P \spcontexte C}
\hspace{10pt} v \in \n,\ C\contextestable
$$
$$
\frac{}{g\envoi{x} \prefix P \spcontexte C \atransition{g\envoi{f(x)}} P \spcontexte C} 
$$
where $C=\contexteqcq$, $x \in \varpile s$ and $x \in \dom f$
$$
\frac{}{g\envoi x\prefix P \spcontexte C \atransition{g\envoi x} P \spcontexte{C'}}
$$
where
\begin{itemize}
\item $C = \contexteqcq$,
$C' = \contexte{\rmpile s {\ensemble x}}{\codomsoustq{q}{\ensemble x}}
{\traceout{\ensemble x}{q}{\rho}}{f}$
\item $x\in\varpile s$ and $x\in q$
\end{itemize}
$$
\frac{}{g \recep x\prefix P \spcontexte C \atransition{g\recep v} P\spcontexte C'}
$$
where
\begin{itemize}
\item  $C=\contexteqcq$, $C'=\contexte{s}{q}{\rho}{f\cup\ensemble{x\mapsto v}}$
\item $x \in \varpile s$, $x$ of type $\nattype$, $v\in \n$
\end{itemize}
$$
\frac{}{g \recep x\prefix P \spcontexte C \atransition{g\recep x} P\spcontexte C'}
$$
where
\begin{itemize}
\item $C=\contexteqcq$, $C'=\contexte{s}{x.q}{\nu\otimes\rho}{f}$
\item $x \in \varpile s$, $x$ of type $\qubittype$, $x \not\in q$
\item $\nu$ density matrix of dimension $2$
\end{itemize}
$$
\frac{}{U[x_1,\ldots,x_n]\prefix P \spcontexte C \tautransition
P \spcontexte{C'}}
$$
where
\begin{itemize}
\item $C=\contexteqcq$, $C'=\contexte{s}{q}{\rho'}{f}$
\item $U\in \transfu$, $x_1,\ldots,x_n \in \varpile s$, and $x_1,\ldots,x_n \in q$ 
\item $\forall\ i,j \in \ensemble{0,\ldots,n}$ such that $i \neq j$ : $\ x_i \neq x_j$
\item $\rho' = \applip{U}{\rho}$
\end{itemize}

\sautdeligne
$\appli{\ }$ is defined in the following way: if $A$ is a $2^n\times 2^n$ matrix, then
$$\appli{A}: \rho \mapsto\adjoint{\Pi}.(A\otimes I^{\otimes k}).\Pi.\rho.\adjoint{\Pi}.
(\adjoint{A}\otimes I^{\otimes k}).\Pi$$
where $\Pi$ is the permutation matrix which places the $x_i$'s at the head of $q$, and
$k = \tailleseq q - n\ $.

$$
\frac{}{M[x_1,\ldots,x_n]\prefix P \scontexteqcq \tautransition
P\scontexte{s}{q}{\rho}{f}}
$$
with
\begin{itemize}
\item $x_1,\ldots,x_n \in \varpile s$ and $x_1,\ldots,x_n \in q$ 
\item $\forall\ i,j \in \ensemble{0,\ldots,n}$ such that $i \neq j$ : $\ x_i \neq x_j$
\item  $M\in\obs$  with $\sum_{i} \lambda_i P_i$ as spectral decomposition
\item $\rho' = \sum_i \applip{P_i}{\rho}$
\end{itemize}
$$
\frac{}{g\envoi{M[x_1,\ldots,x_n]} \prefix P \spcontexte C \tautransition
\debutbloc g\envoi y\prefix \term \finbloc \seq P\spcontexteprob{p_i}{C_i}}
$$
where
\begin{itemize}
\item $C = \contexteqcq$ (which implies $C\contextestable$)
\item $C_i = \contexte{\pileajoutp s {\ensemble{(y,\nattype)}}}{q}{\rho_i}
{\surcharge f \ensemble{y\mapsto\lambda_i}}$
\item $x_1,\ldots,x_n \in \varpile s$ and $x_1,\ldots,x_n \in q$ 
\item $\forall\ i,j \in \ensemble{0,\ldots,n}$ such that $i \neq j$ : $\ x_i \neq x_j$
\item $y$ is a new variable
\item $M\in\obs$  with $\sum_{i} \lambda_i P_i$ as spectral decomposition
\item $p_i = \trace{\applip{P_i}{\rho}}$, \hspace{10pt} $\rho_i = \frac{1}{p_i}\applip{P_i}{\rho}$
\end{itemize}

\subsubsection*{Probabilistic contexts}
$$
\frac{}{P\spcontexteprob{p_i}{C_i} \ptransition{p_j} P\spcontexte{C_j}}
\mbox{ where } \sum_i p_i = 1
$$

\subsubsection*{Sequential composition}
$$
\frac{P \spcontexte{C} \atransition{\alpha} P' \spcontexte{C'}}
{P \seq Q \spcontexte{C} \atransition{\alpha} P'\seq Q \spcontexte{C'}}
\hspace{10pt} \alpha \neq \delta
$$ 
$$
\frac{P \spcontexte{C} \dtransition P' \spcontexte{C'}}{P \seq Q\spcontexte{C} \tautransition Q \spcontexte{C'}}
$$
\subsubsection*{Parallel composition}

In the rules for parallel composition, $C$, $C_P$ and $C_Q$ are defined as:
\begin{itemize}
\item $C = \contexte{\pileajoutp s {(s_P\para s_Q)}}{q}{\rho}{f}$
\item $C_P = \contexte{s_P\pileconcat s}{q}{\rho}{f}$
\item $C_Q = \contexte{s_Q\pileconcat s}{q}{\rho}{f}$
\end{itemize}

In the definition of $C$, the operator $\para$ permits to build a cactus stack (see paragraph \ref{subsecQuantVar}).
In the cactus stack $\pileajoutp s {(s_P\para s_Q)}$ of the process $P\para Q$, the names in $s$
correspond to variables shared by $P$ and $Q$ whereas the names in $s_P$ (resp. $s_Q$) correspond to variables declared in $P$ (resp. $Q$).
$$
\frac{P \spcontexte{C_P} \atransition{\alpha} P' \spcontexte{C_P'}}
{P \para Q \spcontexte{C} \atransition{\alpha} P' \para Q \spcontexte{C'}}
\hspace{10pt} \alpha \neq \delta              
$$
where
\begin{itemize}
\item If $C_P' =\contexte{s'}{q'}{\rho'}{f'}$ then 
$C' = \contexte{\pileajoutp s {(s_P'\para s_Q)}}{q'}{\rho'}{f'}$
with $s_P'$ such that $s' = s_P'\pileconcat s$ ($P$ can neither add to nor remove variables from $s$)

\item If $C_P' =\contexteprob{p_i}{s_i'}{q_i'}{\rho_i'}{f_i'}$
then $C' = \contexteprob{p_i}{\pileajoutp s {({s_P}_i'\para s_Q)}}{q_i'}{\rho_i'}{f_i'}$
with ${s_P}_i'$ such that $s_i' = {s_P}_i'\pileconcat s$
\end{itemize}

$$
\frac{P \spcontexte{C_P} \atransition{g\envoi v} P' \spcontexte{C_P'}
\hspace{15pt}
Q \spcontexte{C_Q} \atransition{g\recep v} Q' \spcontexte{C_Q'}}
{P \para Q \spcontexte{C} \tautransition P' \para Q' \spcontexte{C'}}
$$
where $v \in \n$, $C_Q' = \contexte{s'} {q'}{\rho'}{f'}$, and
$C' = \contexte{\pileajoutp s {(s_P\para s_Q)}} {q}{\rho}{f'}$
$$
\frac{P \spcontexte{C_P} \atransition{g\envoi v} P' \spcontexte{C_P'}
             \hspace{15pt}
              Q \spcontexte{C_Q} \atransition{g\recep x} Q' \spcontexte{C_Q'}}
            {P \para Q \spcontexte{C} \tautransition P' \para Q' \spcontexte{C'}}
$$
where
\begin{itemize}
\item $x \in \varpile s \cup \varpile{s_Q}$, $x$ of type $\qubittype$, $x \not\in q$, $v\in\ensemble{0,1}$
\item $C' = \contexte{\pileajoutp s {(s_P\para s_Q)}} {x.q}{\ket v\bra v\otimes\rho}{f}$
\end{itemize}
$$
\frac{P \spcontexte{C_P} \atransition{g\envoi x} P' \spcontexte{C_P'}
             \hspace{15pt}
              Q \spcontexte{C_Q} \atransition{g\recep y} Q' \spcontexte{C_Q'}}
            {P \para Q \spcontexte{C} \tautransition P' \para Q' \spcontexte{C'}}
$$
where
\begin{itemize}
\item $x \in \varpile s \cup \varpile{s_P}$, $x\in q$
\item $y \in \varpile s \cup \varpile{s_Q}$, $y\not\in q$, $y$ of type $\qubittype$
\item$C' = \contexte{\rmpile{(\pileajoutp s {(s_P\para s_Q)})}{\ensemble x}}
{q[x\leftarrow y]}{\rho}{f}$
\end{itemize}
$$
\frac{P \spcontexte{C_P} \dtransition P' \spcontexte{C_P'}
             \hspace{20pt}
              Q \spcontexte{C_Q} \dtransition Q' \spcontexte{C_Q'}}
            {P \para Q \spcontexte{C} \dtransition \stopproc \spcontexte{C'}}
$$
with
$C' = \contexte{s}{\codomsoustq{q}{e}}{\traceout{e} q \rho}{\domsoust{f}{e}}$ and
$e = (\varpile{s_P}\cup\varpile{s_Q})$

\subsubsection*{Variable declaration}
$$
\frac{}{\debutbloc \declvar{x_1}{t_1},\ldots,\declvar{x_n}{t_n} \findeclvar  P \finbloc \spcontexte{C}
\tautransition \debutbloc P \finbloc \spcontexte{C'}}
$$
with $C = \contexteqcq$, $C' = \contexte{s'}{q}{\rho}{f}$\\
and $s' = \pileajoutp s {\ensemble{(x_1,t_1),\ldots,(x_n,t_n)}}$

\subsubsection*{End of scope of variables}
$$
\frac{P \spcontexte{C} \atransition{\alpha} P' \spcontexte{C'}}
{\debutbloc P \finbloc \spcontexte{C} \atransition{\alpha} \debutbloc P'\finbloc \spcontexte{C'}}
\hspace{10pt}\alpha \neq \delta
$$
$$
\frac{P \spcontexte{C} \dtransition P' \scontexte{\pileajoutp s e}{q}{\rho}{f}}
{\debutbloc P \finbloc \spcontexte{C} \dtransition \stopproc\ 
\scontexte{s}{\codomsoustq{q}{\varpile e}}
{\traceout{\petitevarpile e}{q}{\rho}}{\domsoust{f}{\varpile e}}}
$$

\end{document}